\documentclass{ws-ijmpd}
\usepackage[super,compress]{cite}
\usepackage{url}
\usepackage{siunitx}
\usepackage{hyperref}
\usepackage{xcolor}

\DeclareSIUnit \parsec      {pc}
\DeclareSIUnit \eV                      {eV}
\DeclareSIUnit \keV             {keV}
\DeclareSIUnit \Msun            {M_\odot}

\begin{document}

\markboth{C. R. Arg{\"u}elles et al.}
{structure formation with the quantum nature of dark matter}

\catchline{}{}{}{}{}
%
%%%%%%%%%%%%%%%%%%%%%%%%%%%%%%%%%%%%%%%%%%%%%%%%%%%%%%%%%%%%%%%%%%%%

\title{RESHAPING OUR UNDERSTANDING ON STRUCTURE FORMATION WITH THE QUANTUM NATURE OF THE DARK MATTER%\footnote{For the title, try not to use more than 3 lines.
%Typeset the title in 10~pt Times roman, uppercase and boldface.}  
}

\author{C. R. Arg{\"u}elles,$^{1,2,*}$ E.~A.~Becerra-Vergara,$^{2,3,4}$ A. Krut,$^{2}$ R. Yunis,$^{2,3}$ J.~A.~Rueda,$^{2,3,5,6}$ R.~Ruffini$^{2,3,6}$}

\address{$^{1}$Facultad de Ciencias Astronómicas y Geof\'isicas, Universidad Nacional de La Plata,\\
Paseo del Bosque, B1900FWA La Plata, Argentina\\
$^{2}$International Center for Relativistic Astrophysics Network -- ICRANet,\\
Piazza della Repubblica 10, 65122 Pescara, Italy\\
$^{3}$ICRA, Dipartimento di Fisica, Sapienza Universit\`a di Roma, \\
Piazzale Aldo Moro 5, 00185 Rome, Italy\\
$^{4}$\textit{GIRG}, Escuela de F\'isica, Universidad Industrial de Santander,\\
680002, Cra 27 calle 9, Bucaramanga, Colombia\\
$^{5}$ICRANet-Ferrara, Dip. di Fisica e Scienze della Terra, Universit\`a degli Studi di Ferrara,\\
Via Saragat 1, 44122 Ferrara, Italy\\
$^{6}$INAF, Istituto de Astrofisica e Planetologia Spaziali, \\
Via Fosso del Cavaliere 100, 00133 Rome, Italy\\
$^{*}$carguelles@fcaglp.unlp.edu.ar}

\maketitle

\begin{history}
\received{Day Month Year}
\revised{Day Month Year}
\end{history}

\begin{abstract}
We study the non-linear structure formation in cosmology accounting for the quantum nature of the dark matter (DM) particles in the initial conditions at decoupling, as well as in the relaxation and stability of the DM halos. Differently from cosmological N-body simulations, we use a thermodynamic approach for collisionless systems of self-gravitating fermions in General Relativity, in which the halos reach the steady state by maximizing a coarse-grained entropy. We show the ability of this approach to provide answers to crucial open problems in cosmology, among others: the mass and nature of the DM particle, the formation and nature of supermassive black holes in the early Universe, the nature of the intermediate mass black holes in small halos, and the \textit{core-cusp} problem.
\end{abstract}


\ccode{PACS numbers:}

%\tableofcontents

%%%%%%%%%%%%%%%%%%%%%%%%%%%%%%%%%%%%%%%%%%%%%%%%%%%%%%%%%
%%%%%%%%%%%%%%%%%%%%%%%%%%%%%%%%%%%%%%%%%%%%%%%%%%%%%%%%%
\section{Introduction}
%%%%%%%%%%%%%%%%%%%%%%%%%%%%%%%%%%%%%%%%%%%%%%%%%%%%%%%%%
%%%%%%%%%%%%%%%%%%%%%%%%%%%%%%%%%%%%%%%%%%%%%%%%%%%%%%%%%
%make a discussion about the KNOWN role of the different DM quantum particles at the basis of the seeds of structure formation: state ofthe art with some examples (i.e. how by putting such primordial DM density -either for cold such as WIMPS or for warm such as sterile neutrino candidates- as the source of Einstein equation incosmological perturbation theory brings the corresponding linear (matter) power spectrum, telling how matter will distribute on large scales). But unfortunately, such quantum nature of the particles is not explicitly accounted for in the formation of DM halos within classical N-body simulations (either in LCDM or LWDM paradigms), making us ignorant about their potential effects in the DM halos. (Besides the natural numerical limitations forcing them to use classical point masses of thousands of solar masses or larger)
In the particle DM paradigm, different particle candidates (e.g. WIMPS, axions, sterile neutrinos, etc.) have different early Universe histories that, in turn, translates into different evolution of primordial perturbations in the cosmic plasma which seeds the formation of galaxies at later stages \cite{MoWhite}. The effects of the DM microphysics is, in these cases, often quantified only by the (linear) matter power spectrum: a measure of the distribution of density perturbations before non-linear processes are involved. Such non-linear processes typically include the gravitational collapse of primordial structures, key to the formation of the observed structures in the late Universe, such as galaxies and clusters.

%The microphisics and statistics of the individual particles do not get translated into structure formation
From the point of view of structure formation, the quantum nature of particle DM has been traditionally involved only in the initial conditions before self-gravity comes into play, neglecting any role it may have in the dynamics of the formed systems. 
%However, whether or not this DM microphysics affects the overall structure of DM halos has been recently questioned, since our ignorance about its potential (quantum) effects through the innermost regions of the halos and their stability, represent big holes in our knowledge of these systems. %Original
However, whether or not this DM microphysics affects the overall structure of DM halos has been questioned in recent works. That is, our ignorance about its potential (quantum) effects through the innermost regions of the halos and their stability, may represent a hole in our knowledge of these systems. %RAFA

Indeed, the limited inner spatial resolution of DM halos inherent of cosmological N-body simulations implies huge uncertainties on the DM central distribution. It includes the open questions on how the DM is correlated with baryonic matter on such inner scales, and ultimately, how is the relation with the supermassive black hole (SMBH) at the center of large galaxies, and how they can co-exist in a steady state. Moreover, the use of classical point masses as the building blocks of matter (without considering the quantum nature of the particles) in traditional N-body simulations within the LCDM paradigm does not allow for any new possible sources of quantum-pressure in DM halos (e.g. as recently shown within 3D numerical simulations for bosons~\refcite{Schive} or fermions ~\refcite{Arguelles1}). The arising of these quantum effects at the center of equilibrium systems of self-gravitating quantum particles (either bosons or fermions) is a long-known physical phenomenon \cite{RB}. %ANY equilibrium system?? (Rafa)

Motivated by these open issues, we here present an alternative mechanism of structure formation in cosmology based on thermodynamics of self-gravitating fermions. In such a scenario, the quantum nature of the DM particles is duly taken into account in the physics of decoupling in the early Universe (with corresponding effects in the linear power spectrum), as well as at DM halo formation and subsequent stability and morphology. 
\section{DM halo formation: a thermodynamic approach}
%for collisionless fermions
%%%%%%%%%%%%%%%%%%%%%%%%%%%%%%%%%%%%%%%%%%%%%%%%%%%%%%%%%
%%%%%%%%%%%%%%%%%%%%%%%%%%%%%%%%%%%%%%%%%%%%%%%%%%%%%%%%%

% Insight about violent relaxation from Lynden-Bell
The thermodynamic approach for DM halo formation used here is motivated by the seminal work of \textit{Lynden-Bell} on violent relaxation \cite{L-B}. This approach was originally applied for collisionless stellar systems, and few decades later extended to indistinguishable particles such as massive neutrinos in Ref.~\refcite{Kull}. Relevant extensions to Lynden-Bell theory allowing for out of equilibrium effects (i.e particle evaporation), were performed in Refs.~\refcite{Chavanis1,Chavanis2} using the framework of kinetic theory. Quasi-stationary solutions to such equations (see Eq.~(\ref{Fermionic-Landau}) below), lead to a most-probable, coarse-grained distribution function (DF) at (violent) relaxation, of more realistic applications than the original Lynden-Bell's version: a tidally truncated Fermi–Dirac DF of the form

\begin{equation}
\bar{f}(\epsilon\leq\epsilon_c) = \frac{1-e^{(\epsilon-\epsilon_c)/kT}}{e^{(\epsilon-\mu)/kT}+1}, \qquad \bar{f}(\epsilon>\epsilon_c)=0\, ,
\label{fcDF}
\end{equation}
where $\epsilon=\sqrt{c^2 p^2+m^2 c^4}-mc^2$ is the particle kinetic energy, $\mu$ is the chemical potential with the particle rest-energy subtracted off, $T$ is the effective temperature, $k$ is the Boltzmann constant, $c$ is the speed of light, and $m$ is the DM fermion mass. The full set of dimensionless-parameters of the model are defined by the temperature, degeneracy and cutoff parameters, $\beta=k T/(m c^2)$, $\theta=\mu/(k T)$ and $W=\epsilon_c/(k T)$, respectively (when these parameters are evaluated at the center of the configuration, they will be written with a subscript $0$).

DM halos built upon such a fermionic DF in a cosmological framework, naturally lead to finite-sized halos which can be stable, extremely long-lived, and holding key implications to supermassive BH formation from DM as shown here (and detailed in Ref.~\refcite{Arguelles1}). 

In the limiting case of $\epsilon_c\gg \epsilon$, there is no particle escape, and (\ref{fcDF}) takes the form of the traditional (coarse-grained) Fermi-Dirac DF obtained in the original work of Lynden-Bell. Indeed, this DF is a stationary solution of the Vlasov-Poisson (VP) equation governing collisionless systems \cite{Chavanis1}, that is a solution of Eq.~(\ref{Fermionic-Landau}) shown below, with $\textbf{J}=0$, usually called `VP dynamically-stable' solutions. The more realistic ``tidally-truncated'' DF (\ref{fcDF}) is another kind of quasi-stationary solution of the more general \textit{convection-diffusion} kinetic equation (\ref{Fermionic-Landau}), as demonstrated in Refs.~\refcite{Chavanis1,Chavanis3}. This accounts for the `microphysics' of the long-range particle-particle interactions\footnote{For isolated systems made of fermions, such a current can make (\ref{Fermionic-Landau}) to be of the form of a fermionic Landau equation, conserving mass and energy, and continuously increase the Fermi-Dirac entropy functional (i.e. fulfilling H-theorem) \cite{Chavanis3}.} involved in the diffusion current $\textbf{J}$, caused mainly by the rapid fluctuations of the gravitational potential (proper of violent relaxation processes): 
\begin{equation}
\frac{\partial \bar{f}}{\partial t} + \mathbf{v} \frac{\partial \bar{f}}{\partial \mathbf{r}} - \nabla\Phi \frac{\partial \bar{f}}{\partial \mathbf{v}} = - \frac{\partial \mathbf{J}}{\partial \mathbf{v}}.
\label{Fermionic-Landau}
\end{equation}

When the gravitational fluctuations die down at the end of relaxation, the diffusion current approaches zero (i.e. no more evaporation), and the DF (\ref{fcDF}) remains as a stationary solution of the VP equation. Now, these dynamically-stable VP solutions can be easily used to build DM halos in (hydrostatic) equilibrium able to reproduce galaxy rotation curves (as shown in Refs.~\refcite{Arguelles2,Arguelles3}). However, such analysis does not explain neither \textit{how} such a collisionless self-gravitating system reaches the required steady state, nor if they maximize the coarse-grained entropy $S$ of such a fermion-gas, nor if they are just transitional (unstable/unreachable) states. This stems from the fact that VP dynamical stability \textit{does not} necessarily imply thermodynamical stability, therefore some dynamically stable solutions for DM halos in hydrostatic equilibrium can be more likely to occur in Nature than others!

In order to find dynamically and thermodynamically stable configurations of collisionless self-gravitating fermions in General Relativity (GR), able to model realistic DM halos, it is needed to solve the problem of finding solutions that maximize the global entropy $S=\int_0^R s(r) e^{\lambda/2}4\pi r^2 {\rm d}r$ (at fixed total energy and particle number), where $s(r)$ is the entropy density given by the Gibbs-Duhem relation
\begin{equation}
    \label{s(r)}
    s(r)=\frac{P(r)+\rho(r)-\mu(r) n(r)}{T(r)},
\end{equation}
$e^{\lambda/2}4\pi r^2 dr$ is the proper volume, being $e^{\lambda}$ the spatial component of the spherically-symmetric spacetime metric: ${\rm d}s^2 = e^{\nu}c^2 {\rm d}t^2 -e^{\lambda}{\rm d}r^2 -r^2 {\rm d}\Theta^2 -r^2\sin^2\Theta {\rm d}\phi^2$, with ($r$, $\Theta$, $\phi$) the spherical coordinates, and $\nu$ and $\lambda$ depend only on $r$. 

The mass-energy density $\rho(r)$, pressure $P(r)$ and particle number density $n(r)$ are directly obtained as the corresponding integrals of $\bar{f}(p)$ over momentum space (bounded from above by $\epsilon \leq \epsilon_c$) as detailed in Ref.~\refcite{Arguelles1}. This makes a four-parametric equation of state ($\rho(r)$, $P(r)$) depending on ($\beta,\theta,W,m$) according to the dimensionless parameters in (\ref{fcDF}). The stress-energy tensor is the one of a perfect fluid with the above density and pressure, so the self-gravitating configurations of fermions are solutions of the  hydrostatic equilibrium equation, with the local $T(r)$ and $\mu(r)$ fulfilling, respectively, the Tolman and Klein equilibrium relations (see Ref.~\refcite{Arguelles1} for details). We shall analyze which of such solutions are thermodynamically stable and of astrophysical interest regarding realistic DM halos in cosmology. 
%(linking the local $T(r)$ with the 'Temperautre' $T_\infty$ as seen by an observed at infinity through the metric)

The above stability analysis can be solved in an elegant manner via the `Katz criterion' \cite{Katz}. This allows to find the full set of thermodynamically-stable solutions (i.e. maxima of global entropy $S$) along the full series of equilibrium solutions of self-gravitating fermions. This is a powerful and rigorous method which relies only in the derivatives of the caloric curve given by total energy $E=c^2M(R)$ vs. $1/T_\infty$ (like in Fig.~\ref{fig:CC}), and without the need of calculating explicitly the (rather involving) second order entropy variations $\delta^2 S$. This is a commonly used method to calculate the stability of self-gravitating systems (see e.g. Refs.~\refcite{Chavanis3,Arguelles1}). We work in the microcanonical ensemble applied for isolated systems since, as explained in detail in  Ref.~\refcite{Chavanis3}, it is the most appropriate for astrophysical applications. The relevant thermodynamic potential to be extremized in this ensemble is the coarse-grained entropy $S$. 

In order for such a maximization entropy problem to be well defined, the self-gravitating solutions with given particle number $N=\int_0^R n(r) e^{\lambda/2}4\pi r^2 {\rm d}r$ and total energy $E=c^2M(R)$ must be bounded in radius, as occurring when working with the tidally-truncated DF (\ref{fcDF}). We recall this is not the case for the standard Fermi-Dirac DF (i.e coarse-grained solution of the VP equation), which instead leads to unbounded systems. Therefore, our approach is well posed and without the need to invoke unrealistic bounding boxes \cite{Arguelles1}. 
% Otherwise, as the total mass of the system has no upper bound (i.e. diverges), the and the entropy never reaches a maximum [TremaineBook].
%\footnote{
	%Interestingly, by just solving the extremization of entropy at fixed energy and particle number (i.e. solving up to first order variation in $S$) it is possible to derive the Fermi-Dirac DF (\ref{fcDF}) at statistical equilibrium as well as the GR EoS \citep{2019arXiv190810806C}.} 

%%%%%%%%%%%%%%%%%%%%%%%%%%%%%%%%%%%%%%%%%%%%%%%%%%%%%%%%%
%%%%%%%%%%%%%%%%%%%%%%%%%%%%%%%%%%%%%%%%%%%%%%%%%%%%%%%%%
\section{Astrophysical $\&$ Cosmological applications}
%%%%%%%%%%%%%%%%%%%%%%%%%%%%%%%%%%%%%%%%%%%%%%%%%%%%%%%%%
%%%%%%%%%%%%%%%%%%%%%%%%%%%%%%%%%%%%%%%%%%%%%%%%%%%%%%%%%

We use the following method to assure that the fixed boundary conditions --- such as $N$ and corresponding $M(R)$ required to maximize the global entropy $S$ --- do correspond to values of a given fermionic halo in a realistic cosmological framework: \textbf{(i)} calculate the linear matter power spectrum for such fermionic (quantum) candidates; \textbf{(ii)} apply the corresponding Press-Schechter formalism to obtain the virial halo mass, $M_{\rm vir}\equiv M(R\equiv R_{\rm vir})$, with associated redshift $z_{\rm vir}$ for such a particle mass; and then \textbf{(iii)} use the the most probable coarse-grained DF (\ref{fcDF}) obtained at the end of violent relaxation (valid for that redshift), to obtain the full families of DM (equilibrium) profiles in agreement with above virial constraints. Having done this, we can analyze the (thermodynamic) stability of such halos using the Katz criterion in a given cosmological framework, and further check their astrophysical applications. 

This strategy was recently applied in Ref.~\refcite{Arguelles1} within a WDM cosmology for fermion masses of $\mathcal{O}(10)$~keV, and exemplified here for an average DM halo of mass $M_{\rm vir}\approx 5\times 10^{10} M_\odot$, which started to form at $z_{\rm vir}\approx 10$ according to the Press-Schechter formalism in our cosmological setup.  
%(corresponding to a dimensionless $N=..$ as shown in fig. ; see [MNRAS2021] for details).

\begin{figure}[ht!]
	\centering
	\includegraphics[width=\columnwidth]{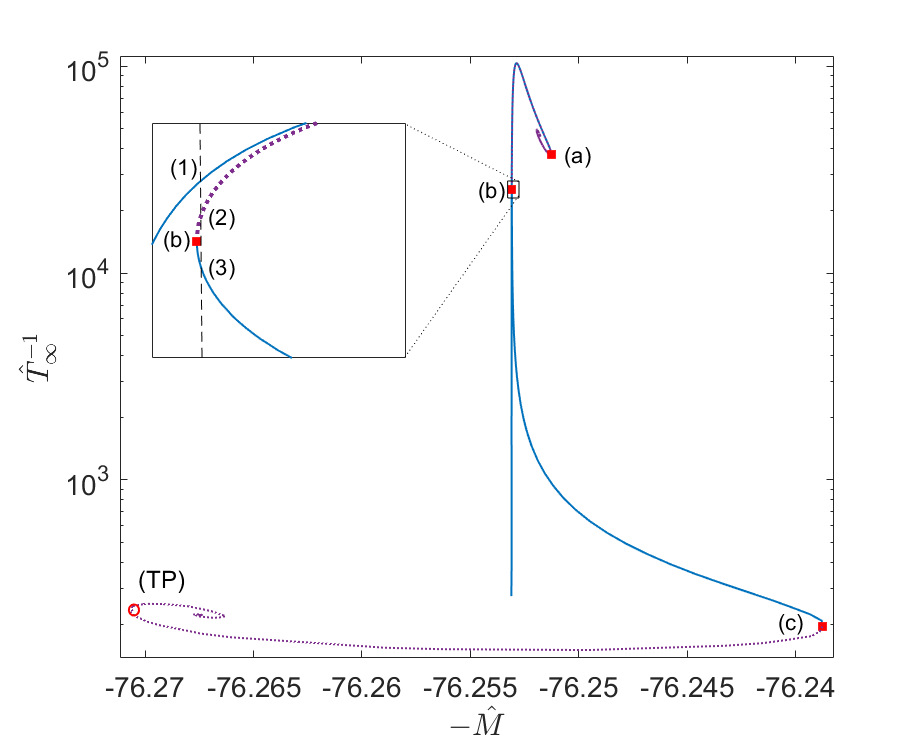}
	\caption{Series of equilibrium solutions along the caloric curve for tidally-truncated configurations of $mc^2=48$~keV fermions, for typical DM halos of $M_{\rm vir}\approx 5\times 10^{10} M_\odot$. The equilibrium states within the continuous-blue branches are thermodynamically (and dynamically) stable, while the dotted-violet branches --- between (a) and (b) and after (c) --- are unstable, according to the Katz criterion. Figure taken from Ref.~\cite{Arguelles1}.}
	\label{fig:CC}
\end{figure}

We show in Fig. \ref{fig:CC} the caloric curve in dimensionless variables, $-\hat{M}$ vs. $1/\hat{T}_\infty$, for an average DM halo (with corresponding $\hat{N}=76.25$, and $W_0-\theta_0=24$ typical of average-size galaxies \cite{Arguelles1}) and for a particle mass of $m=48$ keV. The stability analysis is based on the Katz criterion and follows the global procedure \textbf{(i-iii)} mentioned above. A practical \textit{rule of thumb} to understand how the Katz criterion works is: (a) an unstable mode (coming from stable ones) arises when the caloric curve \textit{rotates clockwise} (i.e. when the negative slope in the ($-\hat{M}$, $\hat{T}_{\infty}^{-1}$) curve becomes infinite for then turning into positive), (b) when the same curve \textit{rotates counterclockwise} a stable mode has been re-gained (the later implying that a positive slope turned into negative just after becoming vertical). In this sense, once in a given unstable branch of the caloric curve, it is necessary as many counterclockwise turns of the curve as clockwise passed, to regain the thermodynamic stability (see Appendix A in Ref.~\refcite{Arguelles1} for details).   

In Fig. \ref{fig:CC}, we display the full family of thermodynamically-stable (i.e. maxima of $S$) solutions in continuous-blue line, and the thermodynamically unstable ones (i.e. minimum or saddle points of $S$) in dotted-violet. Interestingly, there are two spiral features in this caloric curve. The upper one arises due to the fermion (quantum) degeneracy which avoids the  \textit{gravothermal catastrophe} typical of Boltzmannian distributions (as first shown for classical systems in Ref.~\refcite{Chavanis4}). The second spiral feature is of relativistic origin and lead to the turning point criterion for gravitational collapse discussed below.

We now highlight the main conclusions of this stability analysis, and its cosmological and astrophysical consequences in three different subsections.

\subsection{The shape of the fermionic DM halos}\label{sec:A}

%TALK ABOUT the shape of the caloric curve and what it means (including avoidance of gravothermal catstrophe!). 

\begin{figure}[ht!]
	\centering
	\includegraphics[width=\columnwidth]{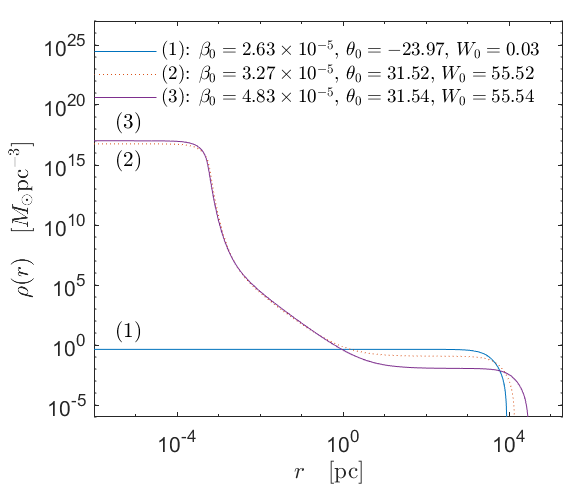}
	\caption{Density profiles for $mc^2= 48$ keV corresponding with three different solutions along the caloric curve in Fig.~\ref{fig:CC} at a fixed energy. Only the profiles (1) (resembling a King distribution) and the \textit{core-halo} one (3) are stable, while profile (2) is thermodynamically unstable. Solutions like (3) are successfully applied to explain the Milky Way rotatoin curve (see Fig.~\ref{fig:vrot}). They are stable, extremely long-lived and fulfill with different observational probes. Figure taken from Ref.~\cite{Arguelles1}.}
	\label{perfk24}
\end{figure}

We analyze from the caloric curve of Fig.~\ref{fig:CC} all the different kinds of density profiles for a conveniently fixed value of the total energy $\hat M$ (see vertical black-dashed line in the inset-box of Fig.~\ref{fig:CC}, leading to the  associated profiles of Fig. \ref{perfk24}). There are two different families of stable and astrophysical DM profiles in a $\mathcal{O}(10)$ keV cosmology, arising from our (thermodynamic) halo-formation approach. They are either King-like, as the solution (1) in Fig. \ref{fig:CC} (resembling the Burkert profile), or they develop a \textit{core-halo} morphology (like the solution (3) in Fig.~\ref{fig:CC}). Importantly, both kinds of solutions can agree with the observational DM surface density relation of galaxies (the latter $\propto \rho(r_{pl})\,r_h$, with $\rho(r_{pl})$ the density at plateau, and $r_h$ the one-halo scale-length of such profiles), as demonstrated in Ref.~\refcite{Arguelles1}. In the first case, the fermions are in the dilute regime (i.e. $\theta_0\ll-1$) and correspond to a global maxima of entropy, while in the second case, the degeneracy pressure (i.e. Pauli principle) is holding the quantum core against gravity, and correspond to a local maxima of entropy. This kind of fermionic \textit{core-halo} solutions are usually referred as the Ruffini-Arg\"uelles-Rueda (RAR) model \cite{jpap4,Arguelles2}. They are extremely long-lived, meta-stable states as demonstrated in Ref.~\refcite{Arguelles1}, with a lifetime $\propto e^{N}$, and more likely to arise in Nature than the former as argued in Ref.~\refcite{Chavanis3}.

\begin{figure}
    \centering
        \includegraphics[width=\hsize,clip]{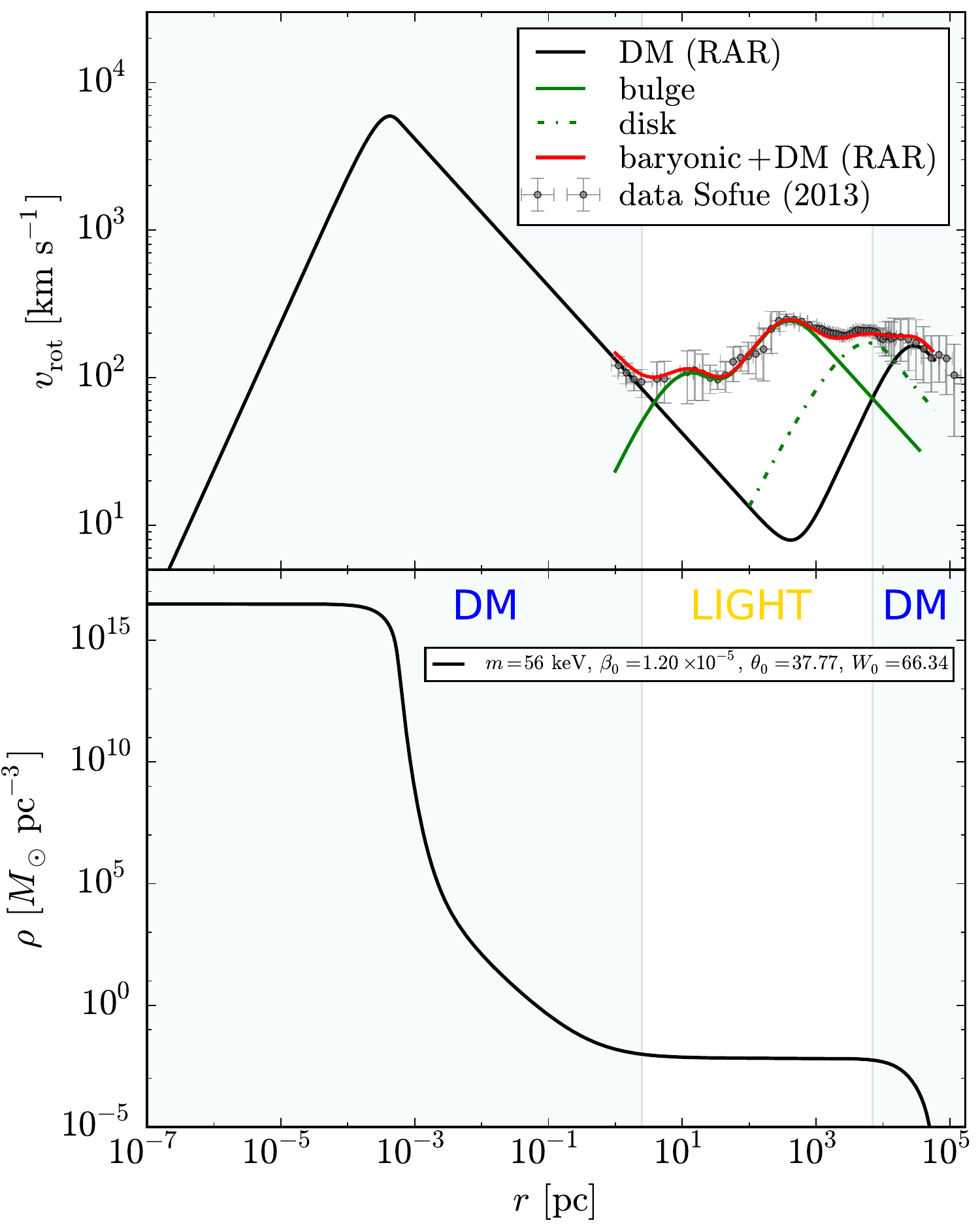}
        \caption{Milky Way rotation curve and DM density profile within the core-halo DM RAR profiles for $mc^2=56$ keV, with a DM quantum core of mass of $M_c = M(r_c)=3.5\times 10^6 M_\odot$. Top: DM (black) and baryonic (bulge + disk) contribution to the rotation curve $v_{\rm rot}$ (total in red). Bottom: corresponding DM density profile. Figure taken from Ref.~\cite{Becerra1}.}\label{fig:vrot}
\end{figure}

\begin{figure}[ht!]%
	\centering%
	\includegraphics[width=0.55\hsize,clip]{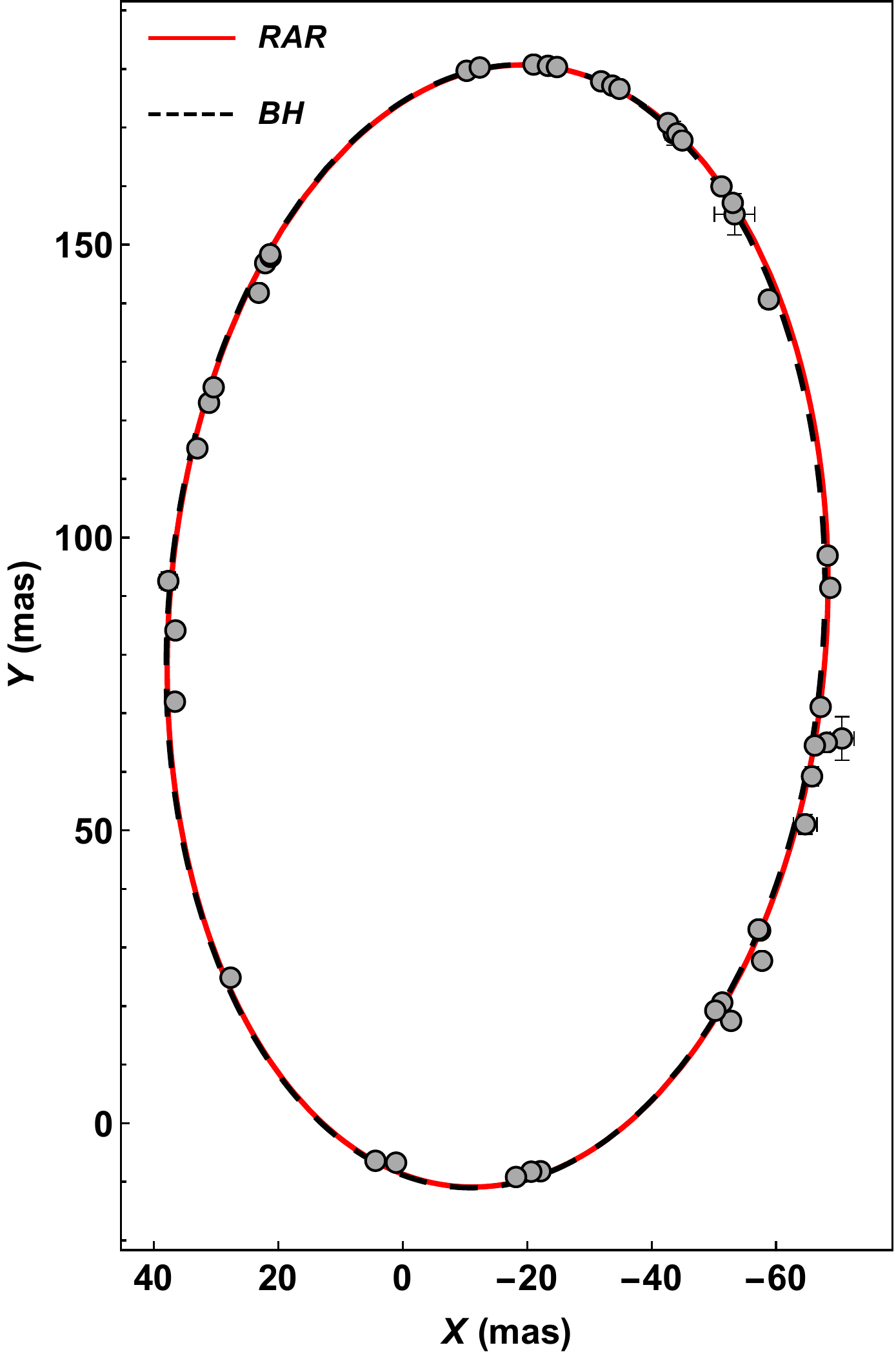}
	\includegraphics[width=0.7\hsize,clip]{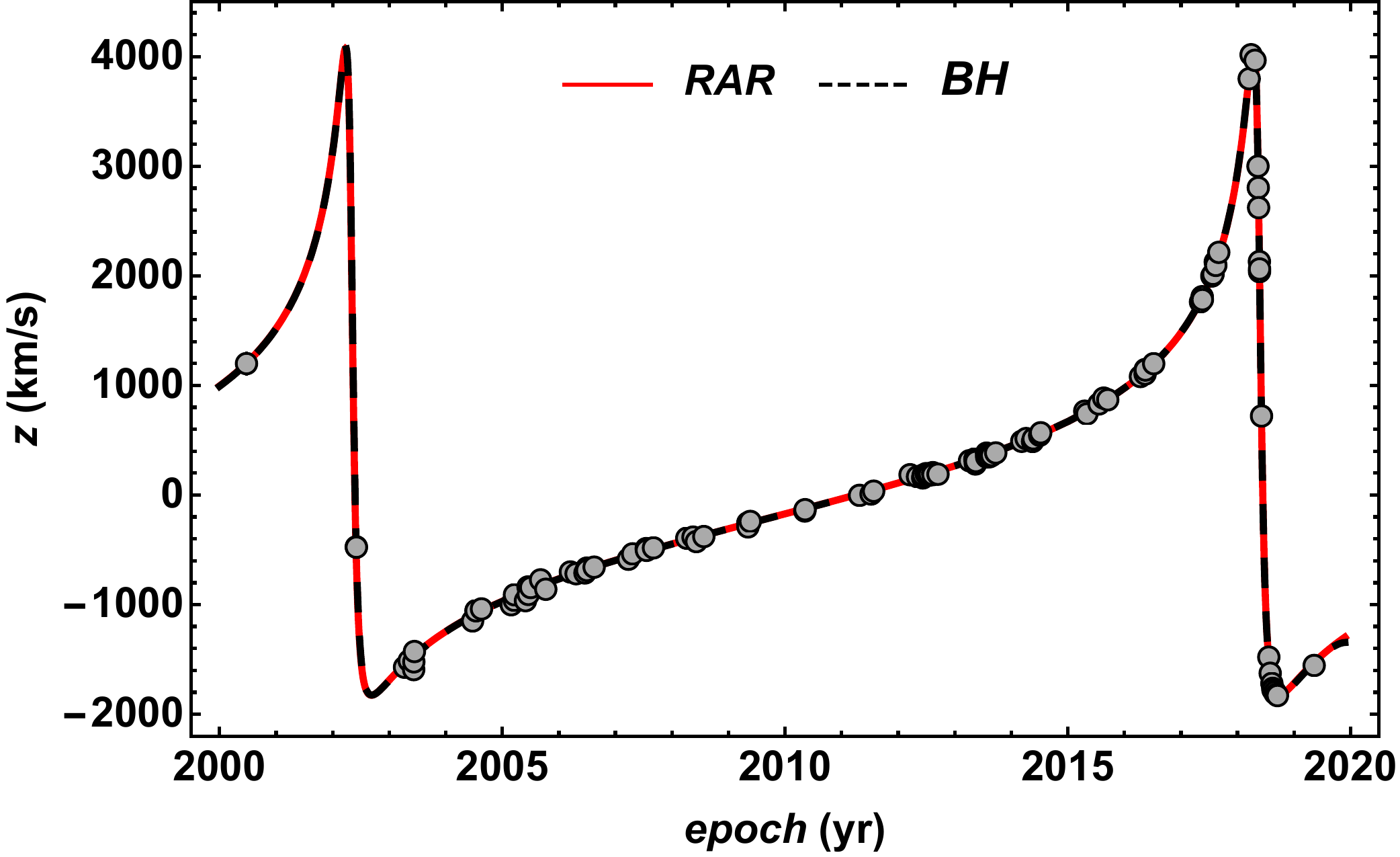}
	\caption{Theoretical and observed orbit and radial velocity ($z$) of S2 around Sgr~A* (wiht data taken from Ref.~\cite{Do}). Top panel: shows the projected orbit on the sky, $X$ (Right Ascension) vs. $Y$ (Declination). Bottom panel: shows the line--of--sight radial velocity as a function of time. The theoretical models are calculated by solving the equations of motion of a test particle in the gravitational field of: (\textit{i}) a Schwarzschild BH of $4.075\times 10^6~M_\odot$ (\textit{black dashed curve}), and (\textit{ii}) the fermionic DM distribution obtained from the core-halo RAR model with $mc^2= 56$ keV fermions (\textit{red curves}). Figure adapted from Ref.~\cite{Becerra1}.}
	\label{fig:S2}%
\end{figure}

\begin{figure}[ht!]% 
	\centering%
	\includegraphics[width=0.75\hsize,clip]{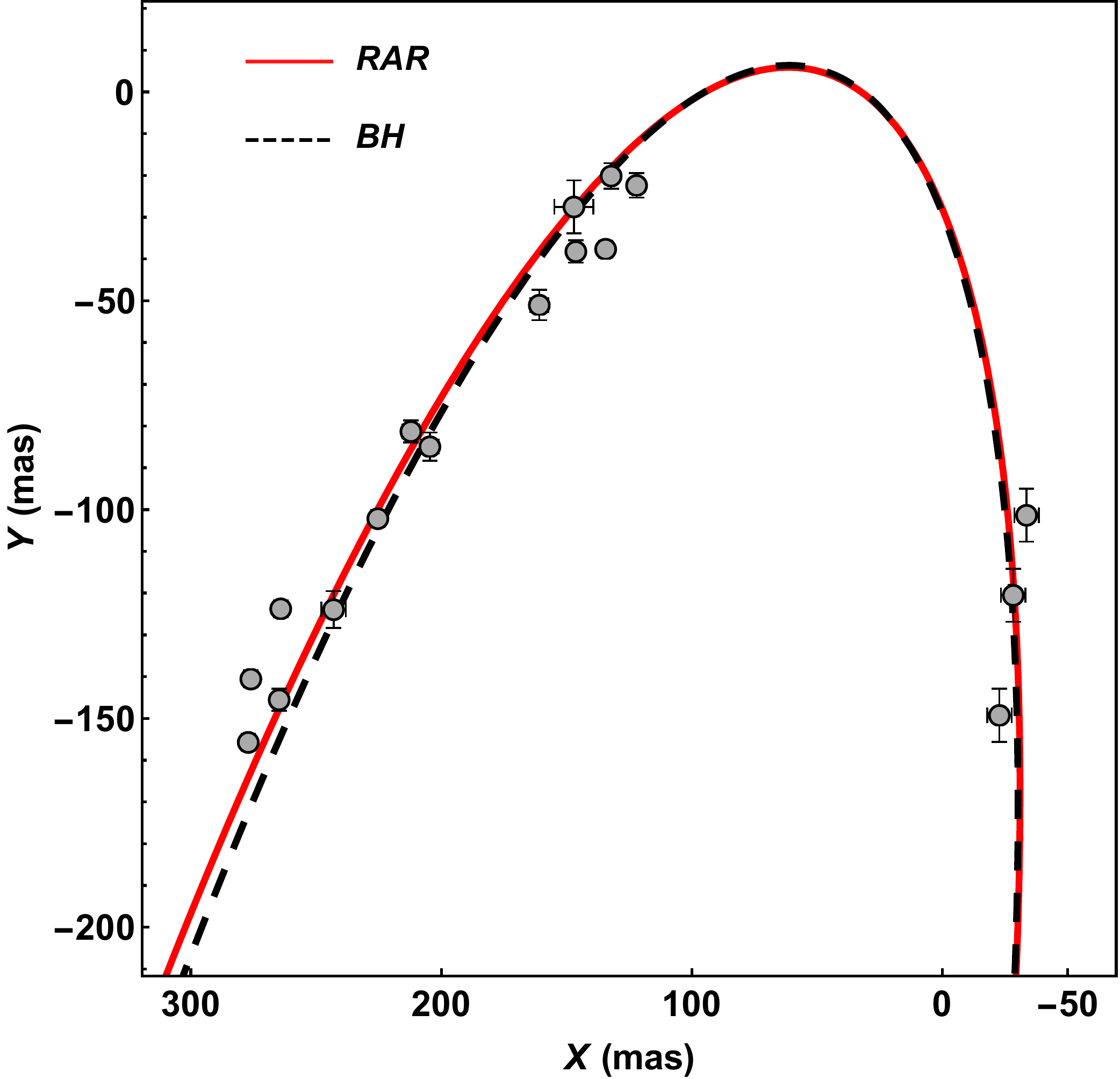}
	\includegraphics[width=0.8\hsize,clip]{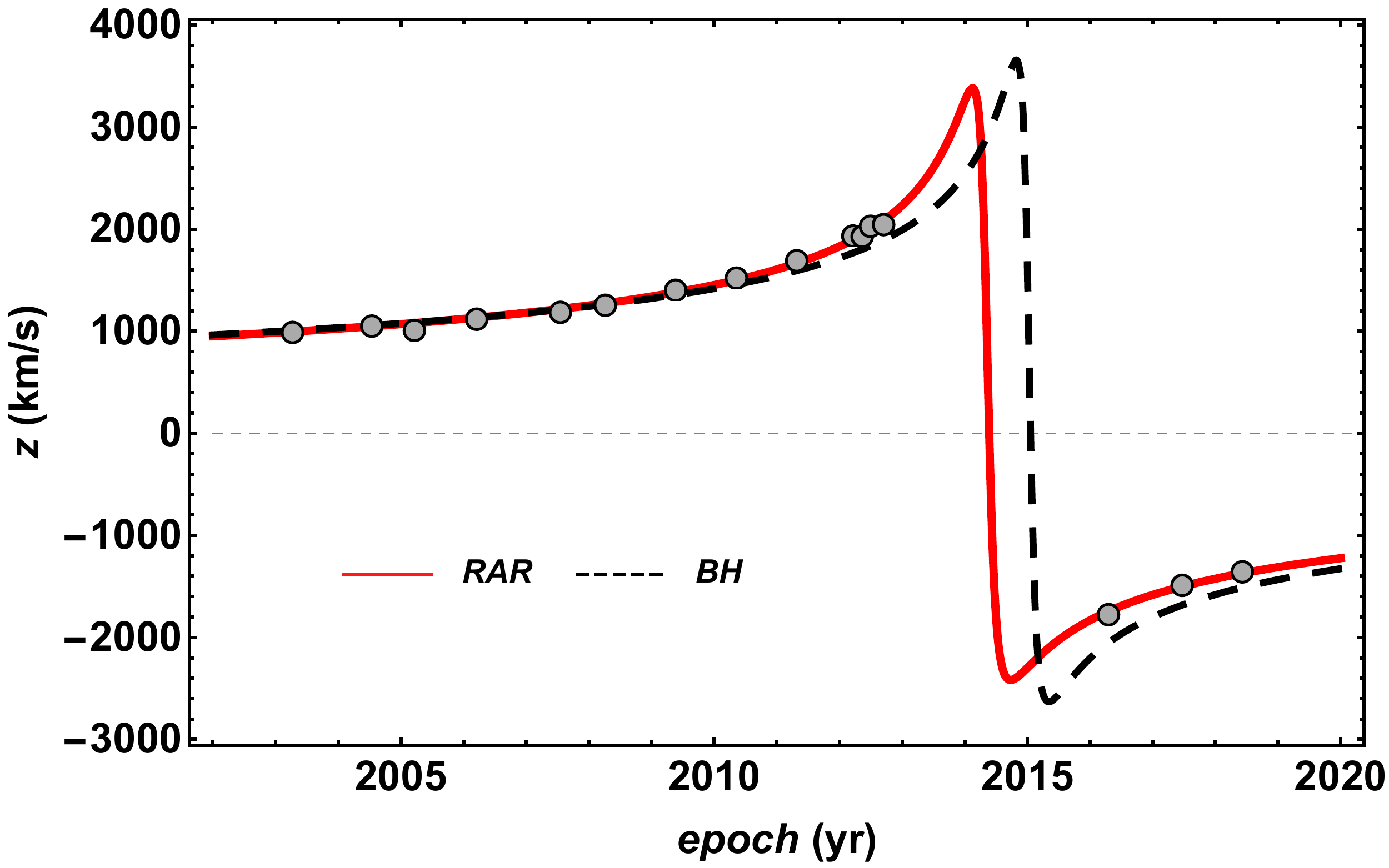}
	\caption{Theoretical and observed orbit and line-of-sight radial velocity of G2 around Sgr~A*. \textit{Top panel}: shows the projected orbit on the sky, $X$ (Right Ascension) vs. $Y$ (Declination). \textit{Bottom panel}: shows the redshift function $z$ as a function of time. The theoretical models are calculated by solving the equations of motion of a test particle in the gravitational field of: (1) a Schwarzschild BH of \SI{4.075E6}{\Msun} (\textit{black dashed curve}), and (2) the DM distribution obtained from the extended RAR model for \SI{56}{\kilo\eV} fermions (\textit{red curves}). The mass of the quantum core in the RAR model {is \SI{3.5E6}{\Msun}}. The observational data are taken from Refs.~\cite{2013ApJ...773L..13P,2017ApJ...840...50P,2019ApJ...871..126G}. Figure adapted from Ref.~\cite{Becerra1}}
    \label{fig:G2}%
\end{figure}

\begin{figure}%
	\centering%
	\includegraphics[width=\hsize]{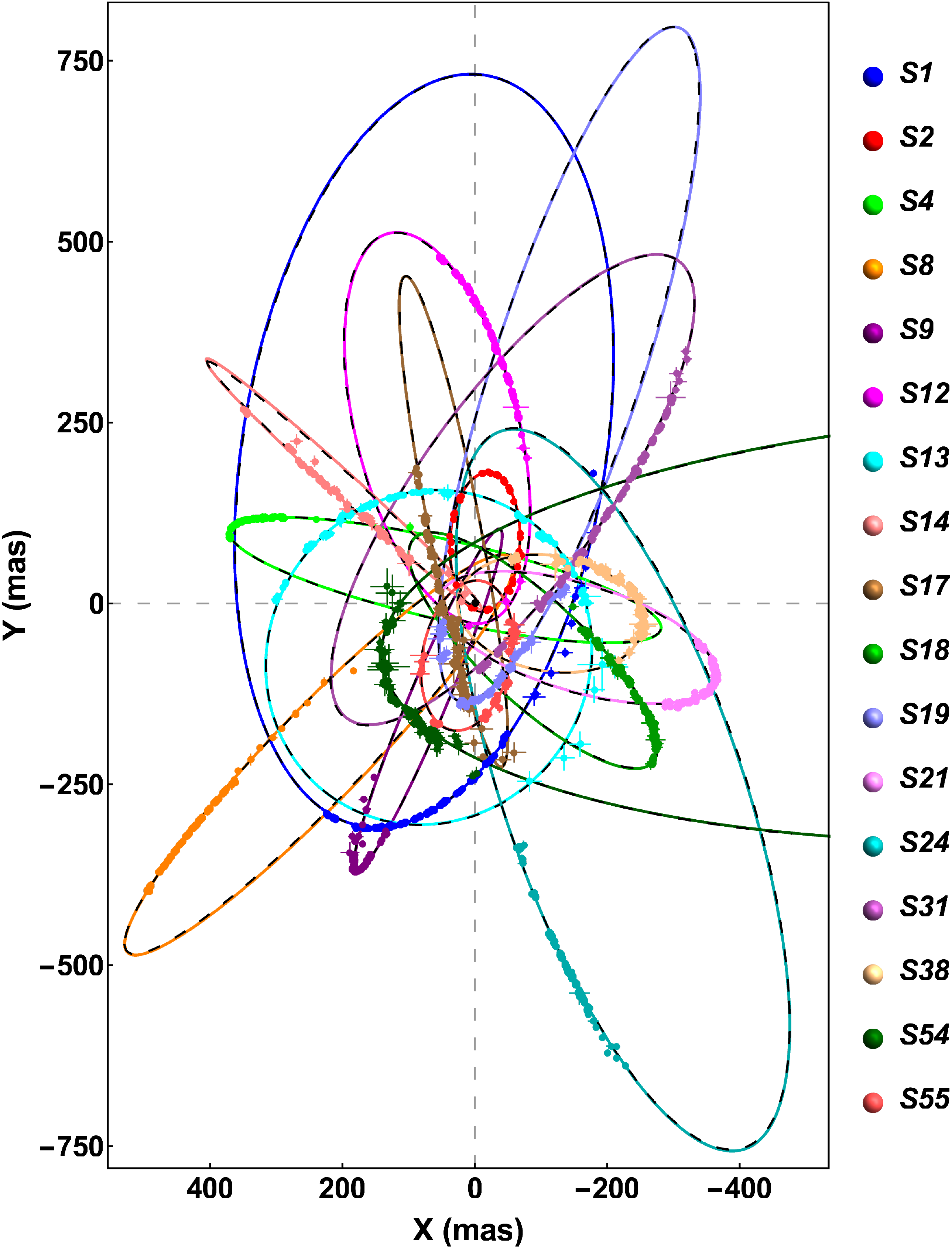}
	\caption{Best-fit orbits for the $17$ best-resolved S-star orbiting Sgr~A*. It shows the projected orbit on the sky, $X$ vs. $Y$, where $X$ is right ascension and $Y$ is declination. The \textit{black dashed curves} correspond to the BH model and the \textit{colored curves} to the RAR model for $m\approx 56$ keV fermions. The astrometric data was taken from Refs.~\cite{Do,Gillessen1,Gillessen2}. Figure taken from Ref.~\cite{Becerra2}.}
	\label{fig:Orbits}%
\end{figure}

Remarkably, solutions like (3) in Fig.~\ref{fig:CC} can fit the rotation curve in galaxies, such as in the case of the Milky Way as shown in Fig.~\ref{fig:vrot}, while their dense quantum core can work as an alternative to the SMBH in non-active galaxies \cite{Arguelles2,Arguelles3,Becerra1,Becerra2}. Indeed, this latter statement was analyzed in the case of our own Galaxy in Ref.~\refcite{Becerra1} within the extended RAR model for $mc^2 = 56$ keV. It has been there shown that the gravitational potential of the DM dense core of such a configuration leads to better fit of the orbits of the S2 and G2 objects around Sgr A*, with respect to the one obtained by the central BH model (see Figs.~\ref{fig:S2} and \ref{fig:G2} ). Furthermore, a geodesic in the gravitational field of the DM profile of the extended RAR model naturally predicts the post-pericenter slowing down phenomenon observed in G2 (see Fig.~\ref{fig:G2}), result that contrasts with the  BH hypothesis where it has been  proposed that G2 is being slowed down by a drag force caused by an accretion flow onto the massive BH over which G2 should move \cite{2017ApJ...840...50P,2019ApJ...871..126G}. Finally, strengthening this DM interpretation of the Galactic Center, in Ref.~\refcite{Becerra2}, this result has been extended to the up-to-date astrometry data of the $17$ best-resolved S-stars orbiting Sgr~A*, achieving to explain the dynamics of the S-stars with similar (and some cases better) accuracy compared to a central BH model (see Fig.~\ref{fig:Orbits}), strengthen the alternative nature of Sgr A*. 

Another key consequence of such a thermodynamic approach for DM halo formation is that, on inner halo scales, the $\mathcal{O}(10)$ keV fermionic density profiles develops an extended plateau (similar to Burkert profiles, with or without the innermost quantum core depending on the degree of central degeneracy), thus not suffering from the core-cusp problem associated to the standard $\Lambda$CDM cosmology \cite{Bullock}. Interestingly, while in N-body simulations (within the CDM paradigm) it is possible to obtain `cored' profiles when including for DM self-interactions (see e.g. Ref.~\refcite{Dave}), here the flatness on inner halo scales are obtained due to the fulfillment of an entropy maximization. The later can be achieved thanks to the long range particle-particle interactions involved in the diffusion-current term of equation (\ref{Fermionic-Landau}) (see also footnote a) as shown here, or complementarily it could be enhanced by (short-range) DM self-interactions as well.

\subsection{Gravitational core-collapse of DM and SMBH formation}

\begin{figure}
	\centering
	\includegraphics[width=\columnwidth]{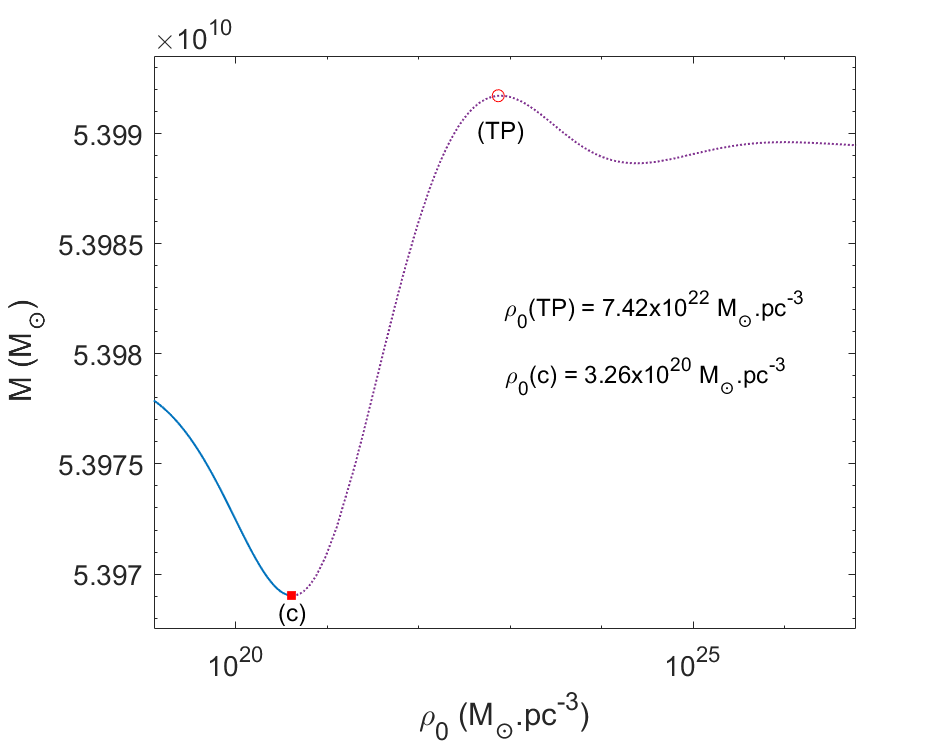}
	\caption{Series of equilibrium states with $N > N_{\rm OV}$ are shown along a $\rho_0$ vs. $M$ curve, in correspondence with the caloric curve of Fig. \ref{fig:CC}. The last stable configuration at the onset of the core-collapse occurs at the minimum of this curve and \textit{prior} to the turning point instability (corresponding with point (c) in Fig. \ref{fig:CC}). Such a critical solution has a critical core mass $M_c^{\rm cr}\approx 2\times 10^8 M_\odot$, and thus a forming a SMBH from DM core-collapse. Figure taken from Ref.~\cite{Arguelles1}.}
	\label{fig:collapse}
\end{figure}

Another striking consequence of the present DM halo formation approach is that it predicts a novel SMBH formation mechanism from DM. Namely, a galactic quantum core made of DM (like in solution (3) of Fig.~\ref{perfk24}) might become so densely packed that, above a threshold critical mass, the quantum pressure can not support it any longer against its own weight, leading to its gravitational collapse and forming a SMBH. Such threshold solution correspond to the last-stable configuration depicted by the point (c) in the caloric curve of Fig.~\ref{fig:CC}, below which there is no possible accessible state (see Ref.~\refcite{Arguelles1} for details). Moreover, such a collapse occurs \textit{prior} to the turning point in the (traditional) \textit{total mass} vs. \textit{central density} diagram (see Fig.~\ref{fig:collapse}). This result, occurring here for self-gravitating systems of neutral fermions at finite $T_\infty$ in GR, agrees with similar findings within rotating perfect-fluid stars \cite{Rezzola}, and generalizes the traditional result applied to compact objects (such as neutron stars) based on full degeneracy ($T\rightarrow 0$).

This kind of DM core-collapse can only occur when the total particle number exceeds the so-called Oppenheimer-Volkoff limit (i.e.~$\hat{N} > \hat N_{\rm OV} \approx 0.4$, as fulfilled in the case here exemplified of $\hat N=76.25$). However, as explained in Sec.~\ref{sec:A}, it can also exist a stable branch of \textit{core-halo} solutions for such $\hat{N} > \hat N_{\rm OV}$ as exemplified through the solution (3) in Fig.~\ref{fig:CC}, before the last stable one (c). This fact is characteristic of self-gravitating solutions of fermions at finite $T$ (as in the RAR model), which differ from the traditional $T\rightarrow 0$ limit where the system must necesarily collapse when $\hat{N} > \hat N_{\rm OV}$ (see Ref.~\refcite{Arguelles1} for further details).

In cases when $\hat{N} > \hat N_{\rm OV}$ as the one shown here, and for $mc^2 \approx 50$ keV, it is possible to form a SMBH of $M_c^{\rm cr}\approx 2\times 10^8 M_\odot$ at the center of a realistic DM halo: that is, the last stable configuration at the onset of the core-collapse (point (c) in Fig. \ref{fig:CC}) correspond to a critical core mass $M_c^{\rm cr}= 2.17\times 10^8 M_\odot$, and thus a forming a SMBH from DM core-collapse (see Ref.~\refcite{Arguelles1} for more details). This is an interesting number, since the majority of the SMBHs are comprised within $\sim 10^8 M_\odot$\cite{Gultekin}, in some way separating active from non-active galaxies. Moreover, the relevance of such a DM core-collapse scenario, is that it can occur within the high $z\sim 10$ Universe when such halos start to form, without the need of prior star formation or any BH seed mechanisms involving (likely unrealistic) super-Eddington accretion rates. This may provide the long-sought solution to the problem of SMBH formation in the early Universe, which deserves further detailed investigations.

Such a critical value triggering core-collapse can be written in a more familiar way in terms of the Planck mass $m_{pl}$ and the fermion mass $m$ as $N_{\rm OV}\approx 0.398\,m_{pl}^3/m^3$, corresponding to a critical (total) mass $M_{\rm OV}\approx 0.384\,m_{pl}^3/m^2$ which for $m\approx 50$ keV yields $M_{\rm OV}\sim 10^8 M_\odot$. Therefore, equilibrium configurations of fermions at finite $T_\infty$ with $M_{\rm vir} < 10^8 M_\odot$ can not undergo core-collapse towards a SMBH within the present scenario. Indeed, the caloric curves in such cases never develop a last-stable solution (like (c) in Fig.~\ref{fig:CC}), nor the second spiral of relativistic origin as shown in Ref.~\refcite{Arguelles1}.

The above numbers imply another intriguing consequence for this new theory: the critical mass for collapse into a massive BH can not be reached for smaller DM halos, for example those surrounding typical dwarf galaxies with $M_{\rm vir} \lesssim 10^8 M_\odot$. Instead, this $\hat{N} < \hat N_{\rm OV}$ scenario would leave smaller dwarf galaxies with a central DM nucleus with a mass typically of $M_c\sim 10^5 M_\odot$ (or less) as shown in Ref.~\refcite{Arguelles3}, thus offering a natural explanation for the so called `intermediate-mass BHs'. Such a DM quantum-core can still mimic the gravitational signatures of a central BH, whilst the dark matter outer halo could also explain the observed galaxy rotation curves \cite{Arguelles3}.

%We proved for the first time that for tidally-truncated self-gravitating systems of neutral fermions at finite $T$ in GR, the thermodynamical (and dynamical) instability occur \textit{prior} to the TP in the $\rho_0$ vs. $M$ curve, as explicited in \cref{sec:TurningPoint}. Indeed, the critical mass of gravitational core-collapse $M_c^{\rm cr}$ is achieved at the last stable configuration (with lower energy with respect to the TP), which interestingly coincides with the maximum but in a \textit{core mass} $M_c$ vs. $\rho_0$ curve. Given the value of $M_c$ at the TP can differ by an order of magnitude below the real $M_c^{\rm cr}$, it shows the importance of this result regarding the SMBH mass estimates, when applied to astrophysics. 

\section{Conclusions}

We have shown that mechanisms of DM halo formation based on entropy maximization in which the quantum (fermionic) nature of the particles is accounted for, may provide answers to crucial open problems in cosmology. In particular they allow: to constraint the mass of the DM particles, to provide a novel mechanisms of supermassive BH formation from DM core-collapse, to understand the nature of the intermediate mass black holes in small halos, and to provide a solution to the \textit{core-cusp} problem.

The approach applied here (and detailed in \cite{Arguelles1}) for DM halo formation in terms of self-gravitating fermions is self-consistent: the nature and mass of the DM particle involved in the linear matter power spectrum calculations (as obtained in Ref.~\refcite{Arguelles1} within a CLASS code for WDM cosmology of $\mathcal{O}(10)$ keV), are the same building blocks at the basis of virialized DM configurations with its inherent effects in the \textit{core-halo} profiles. Importantly, such a fermion mass would produce the same $\Lambda$CDM power-spectrum on scales down to $\mathcal{O}$(Mpc), hence providing the expected large-scale structure (see Appendix B in Ref.~\refcite{Arguelles1}). Moreover, for such typical fermion masses the number of sub-halos in the corresponding WDM cosmology is not in tension with the number of Milky Way satellites \cite{Tollerud}, while being consistent with Lyman-$\alpha$ forest observations \cite{Irsic}. Another relevant consequence is that the maximum entropy approach for structure formation applying here for $\mathcal{O}(10)$ keV particles, can naturally avoid the so called `catch-22' tension arising for fermionic halos within (N-body simulation-based) WDM cosmologies \cite{Maccio}. That is, the resulting fermionic profiles within our approach allow for `cored' density profiles in dwarfs with right sizes and masses, and at the same time in agreement with Lyman-$\alpha$ bounds \cite{Arguelles3}.

To conclude, we believe the results shown here open new insights in the formation and stability of DM halos. Moreover, the quantum degenerate core at the center of the stable DM configurations, and its eventual core-collapse, may play crucial roles in helping to understand the formation of SMBHs in the early Universe, or in mimicking its effects without the need of a singularity at all. The astrophysical consequences of the analysis here shown for $\mathcal{O}(10)$ keV --- together with the recent results in Refs.~\refcite{Arguelles1,Arguelles2,Arguelles3,Becerra1} --- strongly suggest that such DM \textit{core-halo} morphology are a very plausible scenario within the late stages of non-linear structure formation.

\bibliographystyle{ws-ijmpd}
\bibliography{sample}

\end{document}